\documentclass{vgtc}                          
\usepackage[T1]{fontenc}

\graphicspath{{figures/}{pictures/}{images/}{./}} 

\usepackage{times}                     

\usepackage{tabu}                      
\usepackage{booktabs}                  
\usepackage{lipsum}                    
\usepackage{mwe}                       

\usepackage{graphicx} 

\usepackage{mathptmx}                  

\usepackage{enumitem}

\newcommand\changed[1]{{\color{black}#1}}

\onlineid{0}

\vgtccategory{Research}

\vgtcinsertpkg

\title{Grand Challenge: Mediating Between Confirmatory and Exploratory Research Cultures in Health Sciences and Visual Analytics}

\author{Viktor von Wyl\thanks{e-mail: viktor.vonwyl@uzh.ch}\\ %
        \parbox{2.5in}{\scriptsize \centering University of Zurich, Faculty of Medicine \\ UZH Digital Society Initiative (DSI), Zurich}
\and Jürgen Bernard\thanks{e-mail: juergen.bernard@uzh.ch}\\ %
     \parbox{2.5in}{\scriptsize \centering University of Zurich, Department of Informatics  \\ UZH Digital Society Initiative (DSI), Zurich}}

\abstract{

\changed{Collaboration between health science and visual analytics research is often hindered by different, sometimes incompatible approaches to research design.}
Health science often follows hypothesis-driven protocols, registered in advance, and focuses on reproducibility and risk mitigation. 
\changed{Visual analytics, in contrast, relies on iterative data exploration}, prioritizing insight generation and analytic refinement through user interaction. 
These differences create challenges in interdisciplinary projects, including misaligned terminology, unrealistic expectations about data readiness, divergent validation norms, or conflicting explainability requirements. 
\changed{To address these persistent tensions, we identify seven research needs and actions: (1) guidelines for broader community adoption, (2) agreement on quality and validation benchmarks, (3) frameworks for aligning research tasks, (4) integrated workflows combining confirmatory and exploratory stages, (5) tools for harmonizing terminology across disciplines, (6) dedicated bridging roles for transdisciplinary work, and (7) cultural adaptation and mutual recognition. }
\changed{We organize these needs in a framework with three areas: culture, standards, and processes}.
They can constitute a research agenda for developing reliable, reproducible, and clinically relevant data-centric methods.

} 


\begin{document}


\firstsection{Introduction}

\maketitle

Medical and health research is increasingly becoming transdisciplinary. 
The influx of methods from AI into medicine and health sciences \changed{creates great opportunities for new advances and innovation, but also brings new challenges in collaboration~\cite{hicks2019best,jmir2025}}.

\changed{From the clinical perspective, research training emphasizes hypothesis-centered analysis designs, following linear, pre-specified workflows such as pre-registered protocols~\cite{wma2025}.
Data acquisition is treated as a systematic and regulated process to ensure valid results while addressing confounding factors.}
\changed{By contrast, data science and visual analytics (VA) often prioritize iterative, user-centered workflows that refine questions during the exploratory analysis of emerging patterns in pre-existing datasets provided by domain experts.
These approaches assume smooth collaboration and a shared understanding of the data, yet this assumption does not always hold.
}

\changed{
Without effective mediation, differences in methodology, risk tolerance, data governance, and assumptions about data availability can lead to conflicts and misunderstandings.
Even well-intentioned collaborations may fail due to miscommunication, unrealistic expectations, and structural disconnects.
Such difficulties are rooted less in technical incompatibility than in cognitive, procedural, and epistemic mismatches between the domains.
}

\changed{For 15 years, the VAHC workshop on \changed{VA} in healthcare has demonstrated the benefits of combining human expertise with automated analysis for insight generation and data-driven decision-making.
In this grand challenge proposal, we adopt the specific perspective of clinical researchers and health scientists engaged in hypothesis-centered evidence creation, contrast it with the culture of data science and \changed{VA}, identify mediation challenges, and propose needs and actions to address them.
}

\section{Related Challenges}
\label{sec:challenges}

Our experience with transdisciplinary collaborations between health sciences and VA~\cite{von2021challenges,daniore2024wearable,jamia2024,jmir2025} points to \changed{three challenge} areas where frictions may arise: 
\textit{cultures}, \changed{\textit{standards}, and \textit{processes}. 
}

\subsection{Cultures}

\textbf{Explainability and Reproducibility Expectations}: Clinicians \changed{require} transparent reasoning chains with calibrated risk statements tied to clinical thresholds~\cite{ghassemi2021false}. 
\changed{VA} experts may prefer performant, machine-learning-heavy pipelines \changed{that emphasize highlighting visual patterns over statistical certainty}~\cite{heerS12,cashmanHHPDTSMS19}, enriched with explainable AI and trustworthy ML techniques in healthcare~\cite{antweilerF24,lagunaHSCHSCE23}.

\textbf{Risk-Averse vs. Exploratory Culture:} 
\changed{In medicine, strict ethical, legal, and regulatory constraints limit the pace of hypothesis change, as each adjustment must meet compliance and patient safety requirements~\cite{wma2025}.
In \changed{VA}, rapid iteration and exploration are valued to uncover new patterns in interactive frameworks~\cite{zhangGP15}, e.g., for clinical workflow-building or decision support~\cite{wuSLAZ22,prince2024visual,jamia2024}.}

\textbf{Equity, Bias, and Fairness Focus}: 
    Clinicians emphasize equity, bias mitigation, and fairness to ensure care is effective and just across diverse populations, addressing disparities in outcomes, access, and treatment~\cite{wma2025}. 
    \changed{Machine learning-heavy cultures also emphasize preventing biases caused by data processing, machine learning, and visualization to avoid distorting patterns or reinforcing imbalances~\cite{abs-1901-10002,ahnL20,willem2025biases}.}
    
\subsection{Standards}

\textbf{Terminology and Conceptual Gaps}: 
Core terms and concepts such as validation, significance, ground truth, evidence, or knowledge have discipline-specific meanings~\cite{10.1371/journal.pdig.0000347,jmir2025}. 
Without a shared glossary \changed{or agreed definitions, collaborators} may talk past each other~\cite{goldsmith2021emergence,jmir2025}. 

\textbf{Data Governance and Regulatory Compliance}: 
Strict regulations such as GDPR and HIPAA, along with consent management requirements and complex data-sharing agreements in clinical research~\cite{conduah2025data,10.1371/journal.pdig.0000347}, can significantly slow or block the development of \changed{VA} prototypes that rely on open, accessible, and interoperable datasets~\cite{KHALID2023106848}.

\changed{
\textbf{Success Criteria and Benchmarks}: 
Clinical research, particularly randomized controlled trials, optimizes effect size, confidence intervals, and p-values, and requires external, multi-site validation on independent cohorts~\cite{ghassemi2021false,hicks2019best}.
\changed{VA} focuses on insight generation with validations of tool usability and usefulness, sometimes without harmonized benchmarks~\cite{wuCMLBBCDHKKKNP19,bernardSMSPK15,khayatKEG20,prince2024visual}.
}

\begin{figure}[t]
  \centering
  \includegraphics[width=1.0\linewidth]{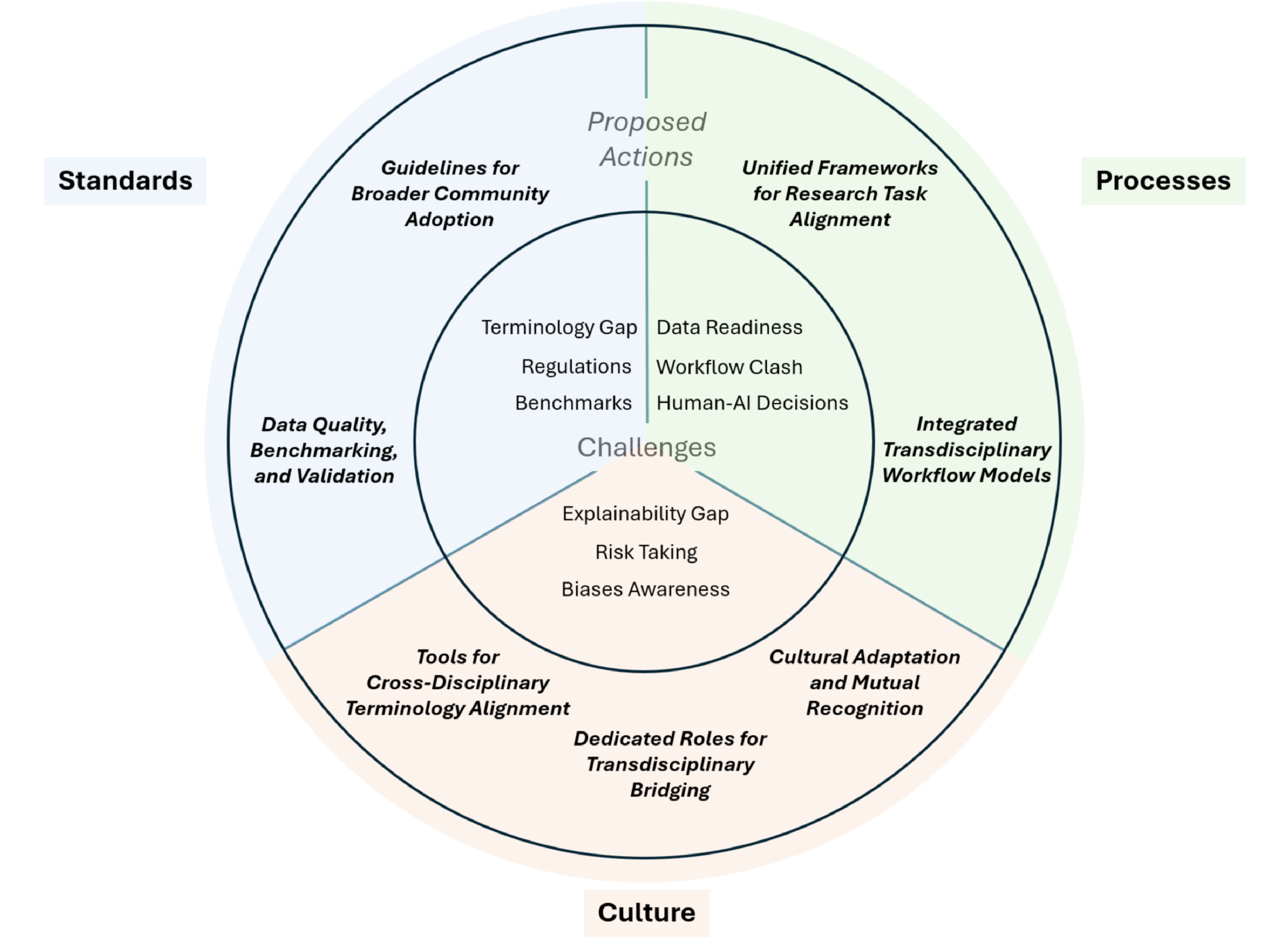}
  \caption{Recurring challenges and overview of seven action areas, structured by three areas: cultures, standards, and processes.}
    \label{fig:framework}
\end{figure}

\subsection{Processes}

\textbf{Data Availability and Readiness}: Health scientists treat data acquisition as a rigorous, \changed{protocol-driven} process, built into study design and governed by protocols and ethics~\cite{wma2025,ghassemi2020review,10.1371/journal.pdig.0000347,hicks2019best}. 
\changed{VA design projects often only begin with existing datasets~\cite{sedlmairMM12}, provided through repositories or domain partners~\cite{RAHMAN2020,abs-2411-07247,abdullahRSGM20}}. 

\textbf{Workflow Philosophy Mismatch}: \changed{Health research optimizes} pre-registered, linear, and hypothesis-centered protocols~\cite{wma2025,jmir2025}, whereas VA refines questions on the fly in exploratory, iterative, and open-ended workflows~\cite{cashmanHHPDTSMS19,tvcg2018Medi,correiaMS25,wuSLAZ22,jamia2024}. 

\textbf{Human–AI Decision Boundaries}: 
Disagreements arise over when decision authority should transfer from the analyst to the algorithm and vice versa, how feedback is recorded and acted upon, and who holds responsibility when blended human–AI decisions fail~\cite{endertHRNFA14,shneiderman20,monadjemiGGGO23,euroVisShortGabi2023,festor2025safety}.
\changed{In health contexts, these boundaries are further complicated by patient safety and liability considerations~\cite{wma2025}.}

\section{Scientific Gaps, Needs, and Proposed Actions}
\label{sec:needs}

Despite a growing recognition of the value of interdisciplinary and transdisciplinary work between health sciences and VA, each new collaboration still often starts from scratch. 
There is \changed{limited availability} of widely accepted guidance, frameworks, or toolkits to scaffold these efforts, \changed{and no broadly accepted best practices to mitigate recurrent misunderstandings, inefficiencies, and cultural differences}.
To address these persistent challenges, we outline required actions, summarized in Figure~\ref{fig:framework}.

\subsection{Cultures}

\subsubsection{Tools for Cross-Disciplinary Terminology Alignment}
There is a need for resources that bridge disciplinary language gaps~\cite{jmir2025}. 
This includes bilingual dictionaries and conceptual concordances that clarify how core terms differ in meaning and use across fields. 
Without these, misunderstandings are frequent and collaboration is slowed. 
Shared lexicons, glossaries, or ontologies can also help establish common ground and prevent semantic drift during project execution.

\subsubsection{Dedicated Roles for Transdisciplinary Bridging}
There seems to be a need for researchers who are explicitly trained to act as mediators between domains. 
These professionals, process facilitators, project integrators, or "bridge-builders", should have cross-domain literacy, cultural sensitivity, and skills in ethical mediation, communication, and collaborative design. 
Their presence can significantly reduce friction and increase project coherence.

\subsubsection{Cultural Adaptation and Mutual Recognition}
Both sides of the collaboration must be willing to adapt culturally. 
\changed{VA} teams need to respect regulatory, ethical, and evidentiary norms from clinical practice, while medical scientists benefit from appreciating the value of iterative modeling and exploratory insights. 
Recognition of transdisciplinary work, including revised \changed{teaching} credit systems, new publication formats, and adjusted evaluation criteria, is necessary to support sustained engagement.

\subsection{Standards}

\subsubsection{Guidelines for Broader Community Adoption}
To scale these efforts, the community must co-develop shared standards that guide collaborative practices, comparable to health research reporting frameworks such as those promoted by the EQUATOR Network~\cite{equator2025}. 
These standards must balance structure with flexibility, offer practical templates and checklists, and gain broad acceptance to become de facto norms in transdisciplinary work.

\subsubsection{Data Quality, Benchmarking, and Validation}
Collaborators need a shared understanding of \changed{data} quality, encompassing data integrity, reproducibility, and external validation~\cite{10.1371/journal.pdig.0000347}. 
Establishing guidelines that bridge the evidentiary expectations of \changed{health science} research with the performance-focused orientation of algorithmic modeling is crucial. 
This includes guidance on benchmarking, multi-site validation, and probabilistic rigor in result interpretation, especially when visualizations are involved.

\subsection{Processes}

\subsubsection{Unified Frameworks for Research Task Alignment}
A shared classification system for research intent, spanning confirmatory, exploratory, and predictive tasks, is essential to align expectations and evaluation criteria early in the process. 
Many frictions stem from differing assumptions about what constitutes valid inquiry or success. 
A framework that maps methodological approaches and their associated metrics (e.g., p-values vs. AUROC) can help avoid such foundational mismatches. 

\subsubsection{Integrated Transdisciplinary Workflow Models}
Successful collaboration demands a hybrid workflow model that reflects the linear, pre-registered rigor of clinical studies and the iterative, question-driven nature of \changed{VA}. 
Such a unified workflow can act as a roadmap to guide planning and coordination, support clearer role definitions, and offer structure without sacrificing flexibility. 
It ensures that both exploratory and confirmatory needs are recognized and integrated coherently.

\section{Conclusion}

In this grand challenge proposal, we \changed{outlined recurring challenges in interdisciplinary and transdisciplinary collaboration between hypothesis-oriented health scientists and \changed{VA} experts, spanning conceptual misalignments, incompatible workflow philosophies, and diverging notions of evidence, explainability, and success.
To address these issues, we proposed scientific needs and actions across the three areas of cultures, standards, and processes.
These needs go beyond technical fixes, requiring formalized standards, aligned workflows, and cultural as well as infrastructural investment to strengthen the maturity, scalability, and long-term sustainability of transdisciplinary practice.
Meeting this challenge will require a coordinated effort to co-design shared infrastructures, train cross-domain mediators, and develop reusable frameworks that enable smoother, more equitable, and scientifically robust collaboration.
Ultimately, this is a call to reframe transdisciplinary research culture, including its methods, tools, languages, values, and reward systems, to bridge the distinct traditions of health sciences and \changed{VA}.}

\acknowledgments{
The authors wish to thank the Digital Society Initiative (DSI) and the DSI Community Health in particular. The DSI is UZH's platform, infrastructure, and interdisciplinary network that shapes the digital transformation of society and science. 
We also wish to thank the members of the \#SOLIDERA working group for inspiration and the UZH Global Fund for funding of the \#SOLIDERA project.}

\bibliographystyle{abbrv-doi}

\bibliography{main}
\end{document}